\documentclass[12pt]{article}
\usepackage[margin=1in]{geometry}
\usepackage{float}
\usepackage{fullpage}
\usepackage{amssymb, amsthm, amsmath}
\usepackage{doublespace}
\usepackage{bm, url}
\usepackage{textcomp}
\usepackage{graphicx}
\usepackage[authoryear]{natbib}
\usepackage{bm}
\usepackage{verbatim}
\usepackage{lineno}
\usepackage{times}
\usepackage{caption}
\usepackage{subcaption}
\usepackage{epstopdf}
\usepackage{float}
\usepackage{bbm}
\usepackage{mathrsfs}

\newcommand{\balpha}{ \mbox{\boldmath $\alpha$}}

\newcommand{\bepsilon}{ \mbox{\boldmath $\epsilon$}}

\newcommand{\boldeta}{ \mbox{\boldmath $\eta$}}

\newcommand{\bA}{ \mbox{\bf A}}

\newcommand{\ba}{ \mbox{\bf a}}

\newcommand{\bX}{ \mbox{\bf X}}

\newcommand{\bZ}{ \mbox{\bf Z}}

\newcommand{\bY}{ \mbox{\bf Y}}

\newcommand{\bh}{ \mbox{\bf h}}

\newcommand{\bs}{ \mbox{\bf s}}

\newcommand{\bH}{ \mbox{\bf H}}
\newcommand{\bS}{ \mbox{\bf S}}

\newcommand{\bW}{ \mbox{\bf W}}

\newcommand{\beq}{ \begin{equation*}}
\newcommand{\eeq}{ \end{equation*}}
\newcommand{\beqn}{ \begin{eqnarray}}
\newcommand{\eeqn}{ \end{eqnarray}}

\begin{document}

\begin{singlespace}
\title{A spatiotemporal recommendation engine for malaria control}


\author{\small{Qian Guan$^{1,}$, Brian J. Reich$^{1}$, and Eric B. Laber$^{1}$}\\ \\
	{\small $^{1}$Department of Statistics, North Carolina State University, Raleigh, North Carolina}\\
}

\date{}
\maketitle

\end{singlespace}

\bigskip
\begin{singlespace}
	\begin{abstract}
		Malaria is an infectious disease affecting a large population across the world, and interventions need to be efficiently applied to reduce the burden of malaria. We develop a framework to help policy-makers decide how to allocate limited resources in realtime for malaria control. We formalize a policy for the resource allocation as a sequence of decisions, one per intervention decision, that map up-to-date disease related information to a resource allocation. An optimal policy must control the spread of the disease while being interpretable and viewed as equitable to stakeholders. We construct an interpretable class of resource allocation policies that can accommodate allocation of resources residing in a continuous domain, and combine a hierarchical Bayesian spatiotemporal model for disease transmission with a policy-search algorithm to estimate an optimal policy for resource allocation within the pre-specified class. The estimated optimal policy under the proposed framework improves the cumulative long-term outcome compared with naive approaches in both simulation experiments and application to malaria interventions in the Democratic Republic of the Congo. 
				\vspace{12pt}
		
	\end{abstract}
\end{singlespace}

\newpage

\section{Introduction}\label{s:intro}
Malaria is a vector-borne infectious disease affecting a large population worldwide, especially in tropical and subtropical regions \citep{guerra2008limits}. It is estimated that there were 216 million cases of malaria resulting in 445,000 deaths globally in 2016 \citep{who2017malaria}, and thus it remains a major public health problem. Effective malaria interventions have been developed, including insecticide-treated mosquito nets (ITNs) \citep{lengeler1998insecticide}, indoor residual spraying (IRS) \citep{pluess2010indoor}, and Artemisinin-based combination therapy (ACT) \citep{eastman2009artemisinin}. However, these interventions are too costly to be given to everyone in need. The malaria research community has made great strides in mapping disease prevalence \cite[][]{bhatt2015effect,bhatt2017improved,kang2018spatio,hay2009world}, modeling its transmission \citep[][]{griffin2010reducing,griffin2014estimates,griffin2016potential,bhadra2011malaria}, developing and testing effective interventions \citep{okell2014contrasting,stuckey2014modeling,parham2015climate,walker2016estimating}, and implementing these interventions in practice. We proposed to build on this work to develop a real-time recommendation engine (RE) for precision interventions to help policy-makers decide how to best allocate limited resources.

Efforts have been made to recommend optimal interventions to different areas to reduce malaria incidence. \cite{okell2014contrasting}, \cite{stuckey2014modeling} and \cite{parham2015climate} conducted cost-effectiveness analysis of different combinations of interventions for malaria control and evaluated how they vary depending on malaria transmission related factors. However they only evaluated the cost-effectiveness of several pre-designed intervention combinations. Also, they either only recommended one most cost-efficient combination of interventions for the whole population in the study area or just recommend a small number of allocation strategies only tailored to environmental factors. They did not provide individualized intervention recommendations that are tailored to small areas or allow these recommendations change over time based on accumulative information. \cite{walker2016estimating} used a dynamic mathematical model to capture the effect of interventions under different malaria transmission settings and recommended intervention choices at national, provincial and pixel level. They only recommended what interventions should be implemented but did not specify the optimal amount of each intervention for each pixel. Also, they did not allow for recommendations to update over time. \cite{bent2017novel} used agent-based exploration techniques to estimate the optimal policy over an analytical policy search space with outcomes predicted by the OpenMalaria Platform \citep{smith2008towards}. They recommended three most cost-effective policies in terms of combination of coverage of ITN and IRS for the specific target population, but generalization of insights to expansive environments is yet to explore and likely to be computationally prohibitive following this approach.

Our proposed recommendation engine (RE) can be formalized as a sequence of decision rules, one per stage of intervention, that map current information to a resource allocation vector specifying the resources allocated to each unit (such as a health zone). The optimal RE should optimize the expectation of some long-term cumulative outcome, e.g. the average malaria prevalence over a given time horizon, and also account for the need for interpretability and fairness. It is related to the idea of dynamic treatment regimes, which are sequential decision rules of giving treatment recommendation to individual patient at each stage based on up-to-date patient information \citep[][]{murphy23o00imalpt,robins2004optimal,chakraborty2013statistical,schulte2014q}. However, there are unique challenges in the spatiotemporal resource allocation problem that distinguish it from typical DTR problems and prevent the direct application of existing DTR methods. First, methods for dynamic treatment regimes typically assume that individuals to receive treatment are independent and a large number of independent trajectories of the individuals are available in the observed data. In our allocation problem, the districts or zones to receive resources are spatially dependent so that we have to treat them as a single allocation objective over the spatial domain at each time point. As a result, there is only single observation available at each time point without independent replications. Second, most of the existing methods for dynamic treatment regimes can only solve problems with discrete treatment space including only a small number of treatment options. However, in our problem the allocation to each region is the proportion of residents to be given a bednet, and so the action space is continuous and high dimensional. \cite{chen2016personalized}, \cite{laber2015tree}, \cite{rich2016optimal} dealt with continuous-valued treatments, but the samples in their methods are assumed independent.

An optimal policy for spatiotemporal resource allocation (e.g., as found using dynamic programming) is a massively complex mathematical function of many geographic and epidemiologic inputs, requiring extensive computation due to the large number of spatial locations under consideration, and is essentially a black box that generates no new knowledge. In contrast, we propose an approach that estimates the optimal RE within an interpretable class of regimes and permits straightforward computation, and is amenable to visualization, scrutiny, and stakeholder input. We build a spatiotemporal model for the progression and transmission of the disease and then draw posterior samples from the fitted system dynamic model to simulate future prevalence for any RE. The mean outcome for each regime within a pre-specified class of interpretable policies is estimated and the optimal RE is maximizer over the class. The idea is related to the policy-search method for dynamic treatment regimes which models the mean outcome as a function of each regime and chooses the maximizer as the optimal regime \citep[][]{robins2008estimation, orellana2010dynamic, zhang2012robust,  zhao2012estimating, zhang2013robust, zhao2015new, zhang2017estimation}. 

However, due to the challenges mentioned above, existing methods cannot be implemented directly to the continuous resource allocation problem. \cite{guan2019Bayesian} used policy-search with g-computation to estimate optimal personalized recommendations for recall intervals for patients with periodontal disease. They built a nonparametric Bayesian model for disease progression process, specified a class of policies on a clinically interpretable risk score, and estimated the optimal policy within this class. However, their observed data included a large number of patients with independent longitudinal profiles, and their decision space was discrete involving only two treatment options. \cite{laber2018optimal} developed a spatiotemporal treatment allocation strategy to slow the spread of white nose syndrome in bats. They used a statistical spatial gravity model to forecast the spread of the disease and developed an algorithm to identify a subset of locations to receive treatment at each time point aiming to optimize long-run control of the disease. While they solved a similar problem, their settings are simpler than what we consider here. Their spatial gravity model is only used to predict the locations that might be affected by the disease at future time points, whereas we need to predict the disease prevalence at each location in the future. Also, they only decided whether or not to apply treatments at each location at each time point so that the decision for each location is binary, whereas we aim to decide the amount of resources to be assigned to each location so that the action space for each location is continuous. Because of these differences, we build a more comprehensive spatiotemporal model to predict disease prevalence and define a new class of policies that can accommodate a continuous action space while remaining interpretable. 

In Section \ref{s:framework}, we formalize the optimal resource allocation problem using a decision theoretic framework. In Section \ref{s:model}, we propose a Bayesian spatiotemporal model to estimate the disease dynamics. In Sections \ref{s:policy} and \ref{s:opt}, we construct a class of resource allocation policies and combine the spatotemporal model with a sequential optimization algorithm to estimate the optimal allocation policy subject to the cost constraint. In Section \ref{s:sim}, we evaluate the performance of the proposed method and compare it with some naive policies using simulation experiments. In Section \ref{s:real}, we apply the proposed method to malaria in the Democratic Republic of the Congo (DRC). Finally, we give a brief discussion in Section \ref{s:discuss}. 

\section{Problem statement}\label{s:framework}
Assume the spatial domain is divided into $n$ health zones. For health zone $l\in\mathcal{L}=\{1,\dots,n\}$, the environmental covariates $(X_{l1},\dots,X_{lp})^T$ are observed at baseline and remain constant over time. At each time point $t=1,\dots,T$, in health zone $l$, the malaria disease rate, $Z_{lt}$, is estimated by the sample proportion of individuals with malaria in the malaria field survey conducted at time $t$. Following \cite{bhatt2015effect}, we use $Y_{lt}=\text{logit}(Z_{lt})$ as the response. For settings with many health zone disease rates near zero, a binomial, Poisson, or negative binomial model might be preferable. The intervention resource allocated to health zone $l$ at time point $t$ is denoted $A_{lt}\in[0,1]$. In our analysis, $A_{lt}$ is the bednet coverage, i.e. the sample proportion of individuals sleeping under a bednet. The data available after time $t$ are $\bH_{t}=\{\bX,\bY_{0},\bA_{1},\bY_{1},\dots,\bA_{t},\bY_{t}\}$, where $\bX=(\bX_1,\dots,\bX_p)$, $\bX_k=(X_{1k},\dots,X_{nk})^T$ for $k=1,\dots,p$, $\bY_t=(Y_{1t},\dots,Y_{nt})^T$ and $\bA_t=(A_{1t},\dots,A_{nt})^T$ represent the collection of covariates, malaria disease rate and resource allocation respectively for all health zones.

We define $\bS_t=\psi_t(\bH_t)\in\mathbb{R}^{q_0}$ to be a summary of the information collected at time $t$, and a resource allocation policy, $\pi$, is a function from $\mathcal{S}=\text{supp }\bS_t$ to the action space $\mathscr{A}=[0,1]^n$. Let $\Pi$ be the class of policies under consideration which we assume are parameterized by $\balpha\in\mathbb{R}^K$. Hence, the policy is determined by the parameter vector $\balpha$ which maps the currently available information to a resource allocation recommendation $\pi(\bS_t;\balpha)$. Define $\Psi(\bs_t)=\Psi(\psi(\bh_t))$ to be the set of possible resource allocation actions with realized history information $\bh_t$ so that $\pi(\bs_t;\balpha)\in\Psi(\bs_t)$.

We formalize the optimal resource allocation policy using potential outcomes. Define $\bZ^\star_t(\overline{\ba}_t)$ to be the potential disease rate at time $t$ under the sequence of allocations $\overline{\ba}_t=\{\ba_1,\dots,\ba_t\}$ up to time $t$. Define $\bZ^\star_t(\pi)$ to be the potential outcome at time $t$ if the resources were allocated under the resource allocation policy $\pi$ up to time $t$. The loss value associated with a policy is defined as the expectation of a loss function of the potential outcomes under the policy, such as the mean malaria prevalence over the next five years (assuming resources are allocated once per year), $L(\pi)=\mathbb{E}\Big\{\big(\sum_{l=1}^n\sum_{t=T+1}^{T+5}Z^\star_{lt}(\pi)\big)/5n\Big\}$. The optimal policy within the pre-specified class is then defined as $\pi_{opt} = \underset{\pi\in\Pi}{\arg\min}\  L(\pi)$. In order to estimate the optimal policy using the observed data, we make the following assumptions \citep[][]{robins2004optimal,schulte2014q}: (1) sequential ignorability, $\{\bZ_{k}^\star(\overline{\ba}_k):\text{for all }\overline{\ba}_k\in\overline{\mathscr{A}_k}\}_{k\geq 1}\perp\!\!\!\perp \bA_t|\bH_t$ for $t=1,\dots$, where $\overline{\mathscr{A}_k}=\mathscr{A}^k$ is the set of all possible resource allocation actions up to time $k$; (2) consistency, $\bZ_t=\bZ^\star(\overline{\bA}_t)$,  where $\overline{\bA}_t$ is the sequence of observed resource allocation actions up to visit $t$; (3) positivity, let $g(\ba_t|\bh_t)$ denote the conditional treatment density given realized history information $\bh_t$, and then there exists $\epsilon>0$ so that $g(\ba_t|\bh_t)\geq\epsilon$ for all $\ba_t\in \Psi(\psi_t(\bh_t))$ and $t=1,2,\dots$. With these assumptions, the loss value function to be optimized can be expressed using g-computation and the data-generating model, which is the malaria transmission model given in Section \ref{s:model}.

\section{Bayesian spatiotemporal model}\label{s:model}
We extend the hierarchical Bayesian spatiotemporal model proposed in \cite{mugglin2002hierarchical} to model the spread of the disease. We assume that
\begin{equation}\label{measure} Y_{lt}=\eta_{lt}+\nu_{lt}\end{equation}
where $\nu_{lt}\overset{iid}{\sim}\text{Normal}(0,\sigma_e^2)$ is measurement error and $\boldeta_{t}=(\eta_{1t},...,\eta_{nt})'$ is a spatiotemporal process. Let $m_l$ be the number of neighbors of zone $l$ and $\mathcal{I}_l$ be the index set of its neighbors. We assume at each time $t$ and each health zone $l$ that
\begin{multline}\label{process_uni}
\eta_{lt}-\eta_{lt-1}=c_0+b_0A_{lt}+(c_1+b_1A_{lt})\eta_{lt-1}+(c_ 2+b_2A_{lt})\Big(\frac{1}{m_l}\sum_{j\in \mathcal{I}_l }\eta_{jt-1}\Big)\\+\sum_{k=1}^{p}\beta_{1k}X_{kl}+\sum_{k=1}^{p}\beta_{2k}X_{kl}A_{lt}+\epsilon_{lt},
\end{multline}
so that the progression of the disease rate depends on its previous disease rate, its neighbors' previous disease rate, environmental covariates and the intervention level.
The coefficients $c_0$, $b_0$, $c_1$, $b_1$, $c_2$, $b_2$ can be interpreted as the intercept, the main effect of resource allocation, the main effect of the previous disease status, the interaction effect of resource allocation and the previous disease status, the main effect of previous neighborhood disease status, the interaction effect of resource allocation and previous neighborhood disease status, respectively. 
The innovation process $\bepsilon_t=(\epsilon_{1t},\dots,\epsilon_{nt})^T$ is modeled as a Gaussian Markov random field (GMRF) that is independent in time and spatially correlated. Specifically, 
\begin{equation}\label{error}\bepsilon_t\sim MVN\{0,\sigma_s^2(M-\rho G)^{-1})\},\end{equation}
where $M$ is diagonal with diagonal elements $m_1,...,m_n$; $G$ is the adjacency matrix with $g_{ll}=0$, $g_{lk}=1$ if zone $k$ is a neighbor of zone $l$, and $g_{lk}=0$ otherwise; $\sigma_s^2$ is the variance parameter, and $\rho\in(0,1)$ controls spatial dependence. An alternative to this model is the intrinsic model with $\rho=1$ which would provide some computational savings at the expense of model flexibility

Equivalently, the spatiotemporal process model can be written in matrix form as the Gaussian first-order autoregressive process:
\beq\label{process_mul}\boldeta_{t} =\bW_t\boldeta_{t-1}+c_{0}+b_0\bA_t+\sum_{k=1}^{p}\beta_{1k}\bX_{k}+\sum_{k=1}^{p}\beta_{2k}(\bX_{k}\circ\bA_{t})+\bepsilon_{t}.\eeq
Spatiotemporal dependence is determined by the $n\times n$ propagator matrix $\bW_t$ that depends on $\bA_t$  and is parameterized by $c_1$, $b_1$, $c_2$ and $b_2$ such that its $(i,j)$ element is
\beq\label{coef}
w_{tij}=
\begin{cases} (1+c_1) + b_1A_{it} &  \text{if }j=i; \\
	(c_2 + b_2A_{it})/m_i &\text{if } j\in \mathcal{I}_i, \text{i.e., zone }j\text{ is the neighbor of zone }i;\\ 0 &\text{otherwise}. \end{cases}
\eeq
The term $c_{0}+b_0\bA_t+\sum_{k=1}^{p}\beta_{1k}\bX_{k}+\sum_{k=1}^{p}\beta_{2k}(\bX_{k}\circ\bA_{t})$ varies depending on the resource allocation $\bA_t$ and environmental covariates $\bX$ to control the latent process $\boldeta_t$ to switch among growth, recession, or stable phases. The latent process is able to incorporate the interaction between the spatial and temporal correlation through the propagator matrix $\bW_t$. Different levels of interventions will affect the environmental effects on the disease rate and also how the disease spreads in space and time. 

Here we assume a first-order autoregressive model in time.  For the motivating malaria analysis this should be sufficient because the time from exposure to manifestation is much less than the yearly time step of the data.  In other settings, higher-order autoregressive models can be fit to account for incubation periods or other delayed effects.

\section{Recommendation engine}\label{s:policy}
We use policy-search to estimate an optimal resource allocation policy within a class of policies that are interpretable to domain experts. We use a global utility function to parameterize the class of policies under consideration. The global utility function $U_G(\ba_t,\bs_t;\balpha)$ summarizes the current available information $\bs_t$ via the parameter $\balpha$ and is a function of the allocation $\ba_t$. We consider the class of policies of the form $\pi(\bs_t;\balpha)=\underset{\ba_{t}}{\arg\max}\text{ }U_G(\ba_t,\bs_t;\balpha)$ subject to some cost constraints, where $\mathcal{U}=\{U_G(\ba_t,\bs_t;\balpha)\}_{\balpha\in\mathbb{R}^K}$ is a class of global utility functions s.t. $U_G(\ba_t,\bs_t;\balpha)$ measures the ``goodness" of allocation $\ba_t$ in state $\bs_t$ under parameter $\balpha$. The class $\mathcal{U}$ is chosen to capture salient features of the decision problem, e.g., cost, fairness, spread dynamics, and logistics. Thus, the resultant class of policies is interpretable and the estimated optimal policy, say indexed by 
$\widehat{\balpha}_{opt}$, can be plugged into this utility function to identify features driving optimal malaria control. In practice, the total available resources are limited by annual logistical or budgetary constraints. For example, constraints are placed on the bednet distribution in each region, $0\leq a_{lt}\leq 1$, and on the total number of bednets distributed so that $\sum_{l=1}^{n}a_{lt}N_{lt}\leq \mathcal{C}(\sum_{l=1}^{n}N_{lt})$, where $N_{lt}$ is the population in health zone $l$ at time $t$.

To construct the global utility function for each health zone $l=1,...,n$, we first define a priority score that represents its priority level of receiving resources so that the zone with higher priority score should be allocated more resources. The priority score is constructed based on all available data before resource allocation at time $t$ and depends on the policy parameter $\balpha$, i.e., $p_{lt}(\bs_{t},\balpha)$. Then we define a local utility $U(a_{lt},p_{lt}(\bs_t,\balpha))$ as a function of $a_{lt}$ that calculates the local utility for health zone $l$ with priority $p_{lt}(\bs_t,\balpha)$ can gain if $a_{lt}$ resources are allocated to the zone. The utility function should be monotonically increasing with the resource allocation since additional resources will bring additional benefits. The global utility function is constructed using the total local utility aggregated over the whole spatial domain.

\subsection{Priority score}
The priority score for each zone $l$ and time $t$ is defined to depend on $q$ user-specified risk factors $\{f_{1lt},\dots,f_{qlt}\}$ constructed from currently available data $\bS_t$. There is great flexibility in constructing the risk factors. They can include the standardized environmental covariates $f_{klt}=X_{jl}$, such as temperature and precipitation, or the zone's current disease status $f_{klt}=Y_{lt}$, the gradient of disease statue $Y_{lt} - Y_{lt-1}$, or even non-linear summaries of the characteristics of the zone. The priority score is then defined as $p_{lt}(\bs_t,\balpha)=1/\{1+\exp[-(f_{1lt}\alpha_1+...+f_{qlt}\alpha_q)]\}\in[0,1]$, where ($\alpha_1,...,\alpha_q$) determine the weights given to each risk factor in the priority score.

\subsection{Local utility function}
Intuitively, a reasonable local utility function $U(a,p)$ that depends on both the resource allocation and the priority score should satisfy the following assumptions:
\begin{enumerate}
	\item $U(0,p) = 0$ for all $p$, i.e. the utility is 0 if no resource is assigned.
	\item $U(a,p)$ is increasing in $a$ for all $p$, i.e., more resources will bring more benefits.
	\item $U(a,p)$ is increasing in $p$ for $0\leq a\leq 1$, i.e., the same level of resources assigned to the health zone with higher priority score will bring more benefits. This property assures the health zone with higher priority score tend to receive more resources in order to maximize the overall utility.  
\end{enumerate}

A natural and simple utility function satisfying the above properties is the linear utility function $U_{lin}(a,p)=
ap$. This utility function assumes that the individual utility increases linearly with resource allocation, i.e., an additional unit of resource will bring constant marginal utility that is set to be the priority of the individual zone.  

Equity or fairness is an important factor to consider when deciding the resource allocation in healthcare \citep[][]{daniels2000benchmarks,gibson2004setting,guindo2012efficacy}. Fairness concerns dictate that everyone should have the same right to the resources and the priority should be given to the zones with more severe situation \citep[][]{nord1999incorporating,nord2015cost}. Similar to the idea proposed in \cite{nord1999incorporating}, a convex utility function can accommodate some fairness concerns. One additional unit of resources will bring more utility if the zone has been assigned less resources.
In other words, the marginal utility value should decrease with the resources allocation. So the utility function that also satisfy the following additional property can account for fairness:

\begin{enumerate}
	\setcounter{enumi}{3}
	\item $\frac{\partial^2}{a}U(a,p)\leq0$ for $0\leq a\leq 1$ and all $p$, i.e., the marginal utility value is decreasing when allocated resource is increasing. 
\end{enumerate}
An example of the utility function that satisfies all above properties is the quadratic utility function $U_{quad}(a,p)=-p(a-1)^2+p$.

\begin{figure}
	\centering
	\caption{Linear (left) and quadratic (right) utility function $U(a,p)$ as a function of resource allocation $a$ for different values of the priority scores, $p$. The linear utility function is $U(a,p)= ap$ and the quadratic utility function is $U(a,p)=-p(a-1)^2+p$ for $0\leq a\leq 1$.}
	\label{f:utility}
	\includegraphics[width= 0.49\textwidth,height=0.4\textwidth]{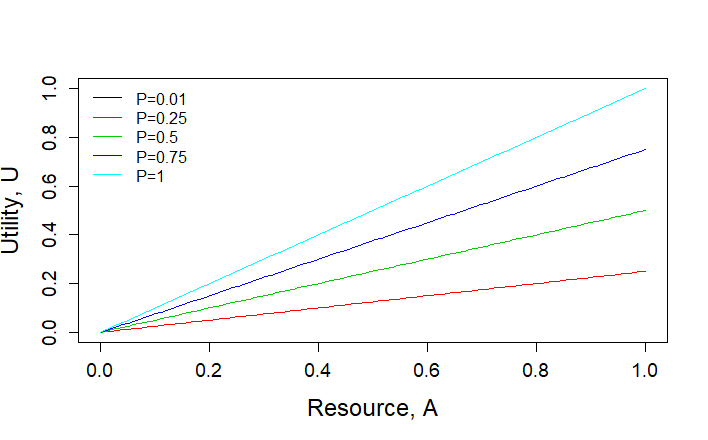}
	\includegraphics[width= 0.49\textwidth,height=0.4\textwidth]{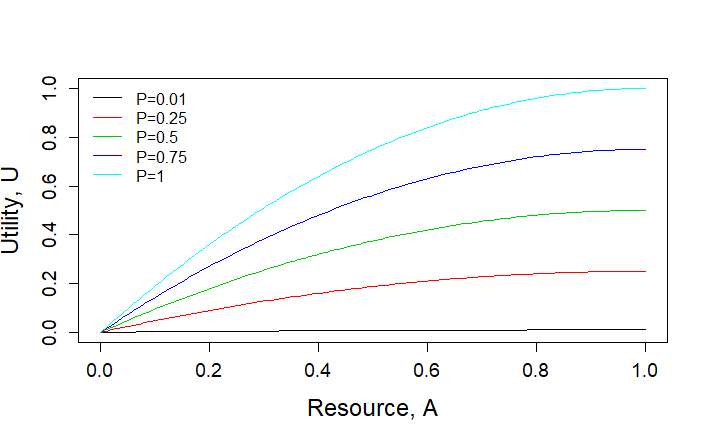}
\end{figure}

Figure \ref{f:utility} illustrates the two utility functions with different values of priority scores. Both utility functions increase with $a$, but the gradient of the linear function in terms of $a$ is constant while the gradient of the quadratic function is decreasing.

\subsection{Global utility function}
The global utility is defined as the summation of all local utilities but also with a penalty on the differences of resource allocations among neighborhood health zones:
\beq\label{global_util}U_G(\ba_t,\bs_t;\balpha)=\sum_{l}U(a_{lt},p_{lt}(\bs_t,\balpha))-\alpha_0\sum_{i\sim j}(a_{it}-a_{jt})^2,\eeq
where $\alpha_0\geq0$, $\balpha=(\alpha_0,\dots,\alpha_q)$, and $i\sim j$ indicates that zone $i$ and zone $j$ are neighbors. The penalty term with a positive weight can smooth the resource allocation and account for fairness. A large penalty $\alpha_0$ encourages spatial clusters of zone to be given intense treatments which could be more effective than scattered sites with intensive treatments. As the weight of the penalty term is to be optimized, this additional term allows for a more flexible class of policies to be considered. 

The recommendation engine suggests allocations that maximize the global utility subject to the resource constraints on the total number of bednets distributed to the whole region:
\begin{equation}\label{constr}\begin{aligned}
\pi(\bs_t;\balpha)=\underset{\ba_{t}}{\arg\max}\text{ }U_G(\ba_t,\bs_t;\balpha)\\
\text{such that } \sum_{l}a_{lt}N_{lt}/(\sum_{l}N_{lt})\leq \mathcal{C}\text{ and }0\leq a_{lt}\leq 1.\end{aligned}\end{equation}

Other nuances in the resource allocation decision making can be easily taken into consideration by adding more constraints to (\ref{constr}). For example, if it is agreed that allocating bednets to health zones with malaria prevalence $<1\%$ is ill-advised, we can simply add a constraint that $a_{lt}=0$ if $z_{lt}<0.01$.

\section{Policy-search}\label{s:opt}
The optimal priority score weights $\balpha$ minimizes an optimality criterion $L(\balpha)$, such as the estimated expected cumulative (over space and time) malaria prevalence over the next five years. Therefore, although the policy only gives the resource allocation of one time point, the policy optimizes long-term outcomes.
Given the posterior samples of the parameters in the Bayesian spatiotemporal model, the future malaria prevalence can be simulated to construct an estimator of the expected cumulative prevalence over space and time $\tilde{L}(\balpha)$. The plug-in estimator of the optimal weights is $\tilde{\balpha}_{opt}=\underset{\balpha}{\arg\min}\text{ }{\tilde{L}(\balpha)}$. For numerical stability, we replace $\tilde{L}(\balpha)$ with $\hat{L}(\balpha)=\tilde{L}(\balpha)+0.0001\sum_{i=0}^{q}\alpha_i^2$. Then $\widehat{\balpha}_{opt}$ is defined as the minimizer of $\hat{L}(\balpha)$.

A Kriging-based optimization method \citep{picheny2013quantile} is used to estimate $\balpha$ which minimizes $\hat{L}(\balpha)$. At the first step, we evaluate the value for the initial $100$ points of $\balpha$ from a Latin hypercube design generated using ``optimumLHS" function in the R package ``LHS." We do a linear transformation of the initial design by setting the range of $\alpha_1,\dots,\alpha_q\in[-5,5]$ and $\alpha_0\in[0,1]$. The value $\hat{L}_k$ corresponding to each point $\balpha_k$ in the design is estimated using Monte Carlo simulation given the posterior samples of the parameters. By using a different posterior sample of the model parameters for each simulated trajectory in the Monte Carlo simulation, trajectories are samples from the full posterior predictive distribution of future prevalence and thus our policy accounts for both parametric and aleatoric uncertainty. Using the initial training data $\{\balpha_k,\hat{L}_k\}$, a Gaussian process regression model is fit to predict the value corresponding to a new $\balpha$ using kriging. The algorithm sequentially selects next weight $\balpha$ to visit that optimizes the expected improvement (EI) defined in \cite{jones1998efficient} and updates the Gaussian process model parameters at each iteration. We use the the R package ``DiceOptim" \citep{roustant2012dicekriging} to implement the optimization procedure.

At each of the future years, if new data is available, the Bayesian spatiotemporal model can be refitted and the posterior distribution of the parameters can be updated. The resource allocation decision for the future years can be made by reoptimizing the priority score weights based on the updated posterior samples.

Each posterior sample of the parameters and predicted malaria prevalence in the MCMC procedure corresponds to a different optimal $\balpha$. We can quantify the uncertainty of $\hat{\balpha}_{opt}$ by applying the optimization procedure for each posterior draw to get the posterior distribution for $\balpha_{opt}$.
\section{Simulation}\label{s:sim}

\subsection{Generative model}

For the simulation, we assume that there are $n=100$ health zones that are arranged as a $10\times10$ square grid with grid spacing 1 between adjacent sites, and include only one environmental covariate $X_{l1}$ simulated from the Gaussian process with mean zero and variance one, and correlation $Cor(X_{l1},X_{j1})=\exp(-d_{lj}/2)$, where $d_{lj}$ is the distance between the centroids of zone $l$ and zone $j$. We simulate the baseline latent process for the health zones from a Gaussian process such that $\boldeta_0=(\eta_{10},\dots,\eta_{n0})^T\sim MVN(0,0.5\label{key}^2(M-0.9G)^{-1})$ and the corresponding logit transformation of the disease rates are simulated from $Y_{l0}\sim N(\eta_{l0},0.01^2)$ for $l=1,\dots,n$. We simulate the disease spread for $T=5$ years following the generative model such that for $l=1,\dots,100$, $t=1,\dots,5$,
\begin{multline*}
\eta_{lt}=(0.9-0.1A_{lt})\eta_{lt-1}+(0.1-0.1A_{lt})/m_l\sum_{j\in N_l }\eta_{jt-1}\\+0.2-0.7A_{lt}+0.12X_{1l}-0.1X_{1l}A_{lt}+\epsilon_{lt},
\end{multline*}
where $\bepsilon_t\sim MVN(0,0.1^2(M-0.9 G)^{-1})$, and resource allocation is assumed increasing over time (to mimic the real malaria data) such that $A_{lt}$ are simulated from $N(0.1*t,0.05^2)$ and are truncated at $0$ and $1$. The disease rates for the five years are simulated from $Z_{lt}\sim N(Y_{lt},0.01^2)$ for $l=1,\dots,n$ and $t=1,\dots,5$. We simulate $100$ datasets using this generative model. 

%
\subsection{Policy estimation}
We consider three risk factors in the policy: the environmental covariate $f_{1lt}=X_{l1}$, logit of disease rate at previous time point $f_{2lt}=Y_{lt-1}$, mean logit of disease rates of neighborhood zones at previous time point $f_{3lt}=\sum_{j\sim l}Y_{jt-1}/m_l$. The priority score of zone $l$ at time $t$ is then $1/\{1+\exp[-(\alpha_1X_{l1}+\alpha_2 Y_{lt-1}+\alpha_3\sum_{j\sim l}Y_{jt-1}/m_l)]\}$. We assume that the number of individuals in different health zones are the same and give the constraint that $\frac{1}{n}\sum_lA_{lt}\leq \mathcal{C}$. We consider three scenarios with different values of resource constraint level $\mathcal{C}=0.2,0.5$ and $0.8$.


We consider the following resource allocation policies for comparison:
\begin{enumerate}
	\item Linear utility (Linear): our proposed policy with linear local utility function and a spatial penalty term to smooth the resource allocation.
	
	\item Quadratic utility (Quad): our proposed policy with quadratic local utility function and a spatial penalty term to smooth the resource allocation.
	
	\item Highest rate (Highest\_rate): assign a bednet to each individual in the $n\mathcal{C}$ zones with highest disease rates and no bednets to the remaining zones.
	\item Even: assign the same percentage of bednets $\mathcal{C}$ to each healthzone.
	
\end{enumerate}

For each simulated dataset, we use all information simulated up to year $T=5$ as the training data to fit our proposed Bayesian spatiotemporal model using MCMC sampling with $5,000$ iterations. For the first two policies, the optimality criterion $L(\balpha)$ for each $\balpha$ is estimated using Monte Carlo simulations given the posterior samples. Sequential optimization is used to estimate the optimal policy that minimizes the expected mean malaria prevalence in the future five years within the pre-specified class. The loss value associated with estimated policy are approximated by $1,000$ Monte Carlo simulations given the true generative model. For the last two policies, there is no need to fit the model and no parameters to estimate. We just use $1,000$ Monte Carlo simulations given the true generative model to approximate the expected mean malaria prevalence in the future five years under the policy. The approximated loss values associated with the four policies for $i$th simulated dataset are denote as $L_{l}^i,L_{q}^i,L_{hr}^i,L_{ev}^i$ respectively.

\begin{figure}
	\caption{\small{The improvement of the proposed policies with linear utility function or quadratic utility function compared with 		``Highest\_rate" policy (left) and ``Even" policy (right) with different resource constraints $\mathcal{C}=0.2,0.5$ and $0.8$ when the model is correctly specified (upper) or misspecified (lower).}}\label{f:imprv}
	\begin{subfigure}{1\textwidth}
		\label{f:imprv_notmis}
		\caption{Model is correctly specified}
		\begin{center}\begin{picture}(420,220)
			\includegraphics[height=0.48\textwidth]{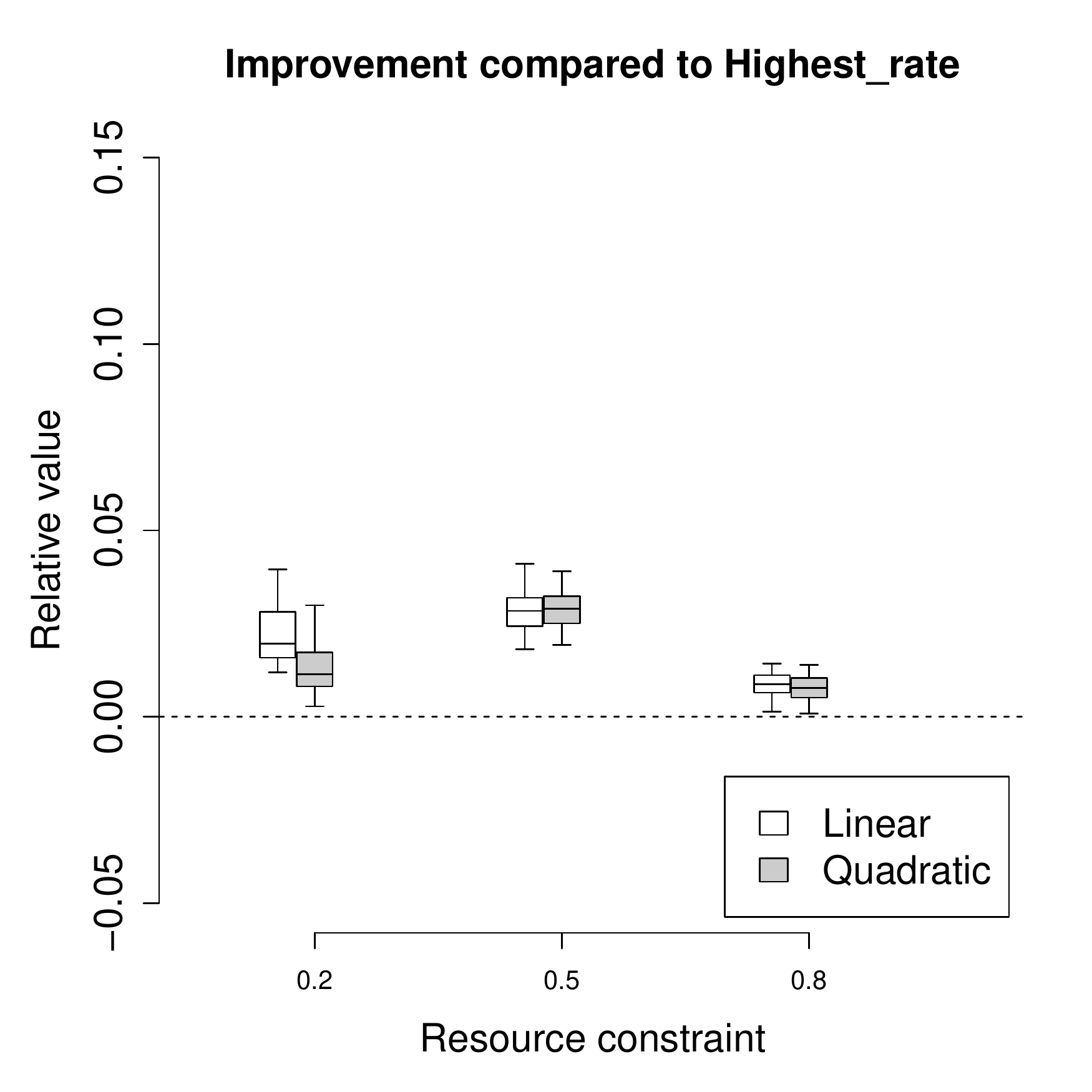}
			\includegraphics[height=0.48\textwidth]{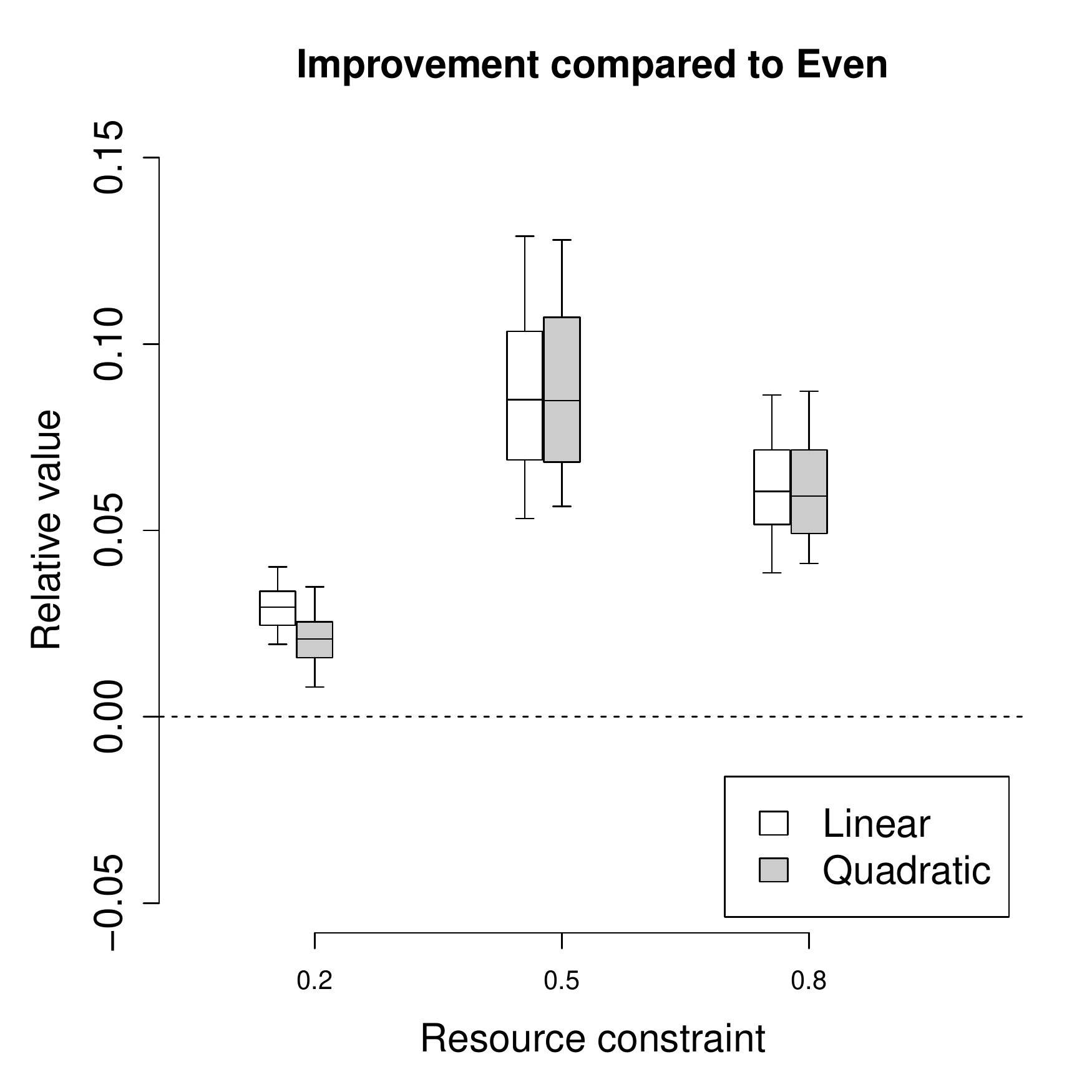}
			\end{picture}
		\end{center}
	\end{subfigure}
	\begin{subfigure}{1\textwidth}
		\label{f:imprv_mis}
		\caption{Model is misspecified}
		\begin{center}\begin{picture}(420,220)
			\includegraphics[height=0.48\textwidth]{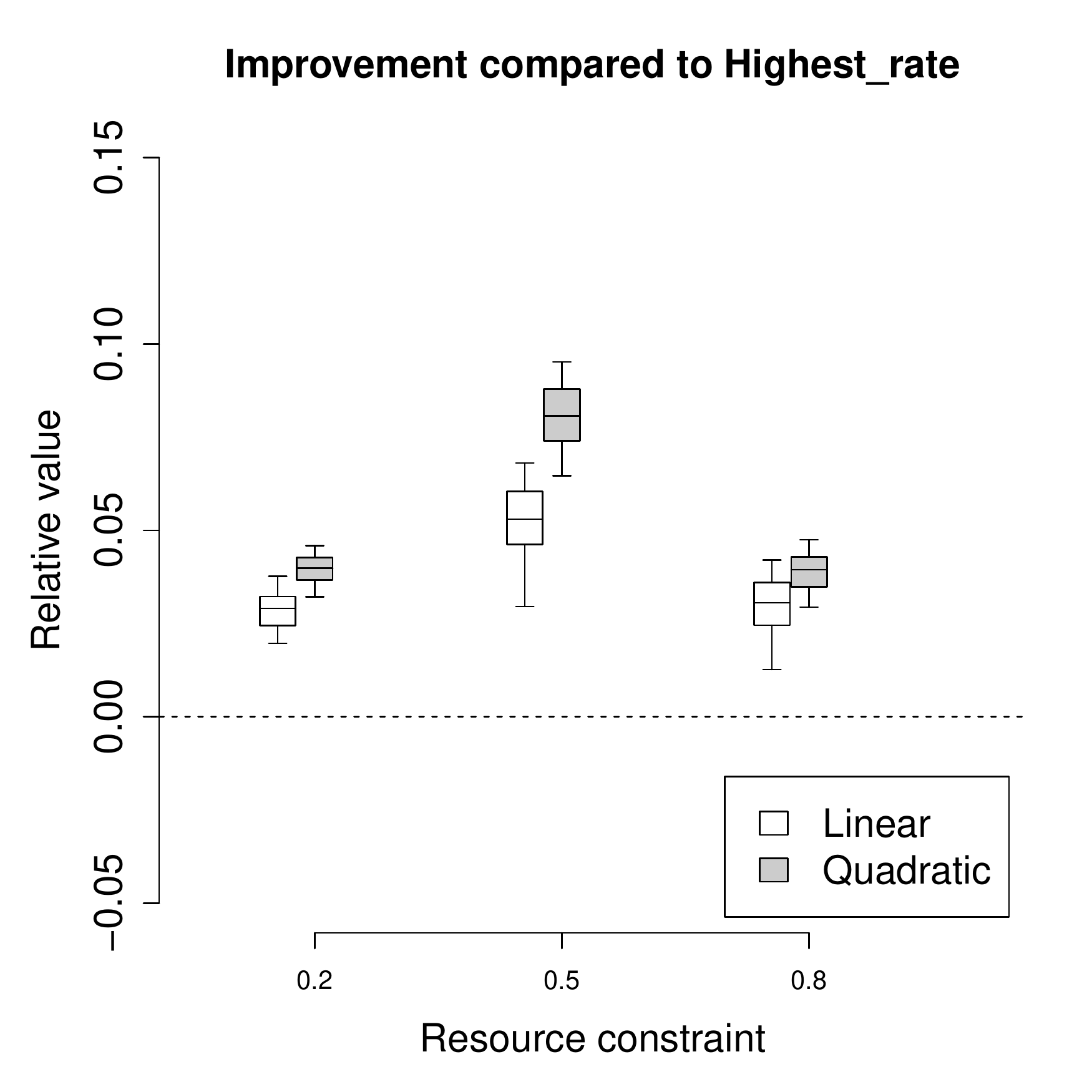}
			\includegraphics[height=0.48\textwidth]{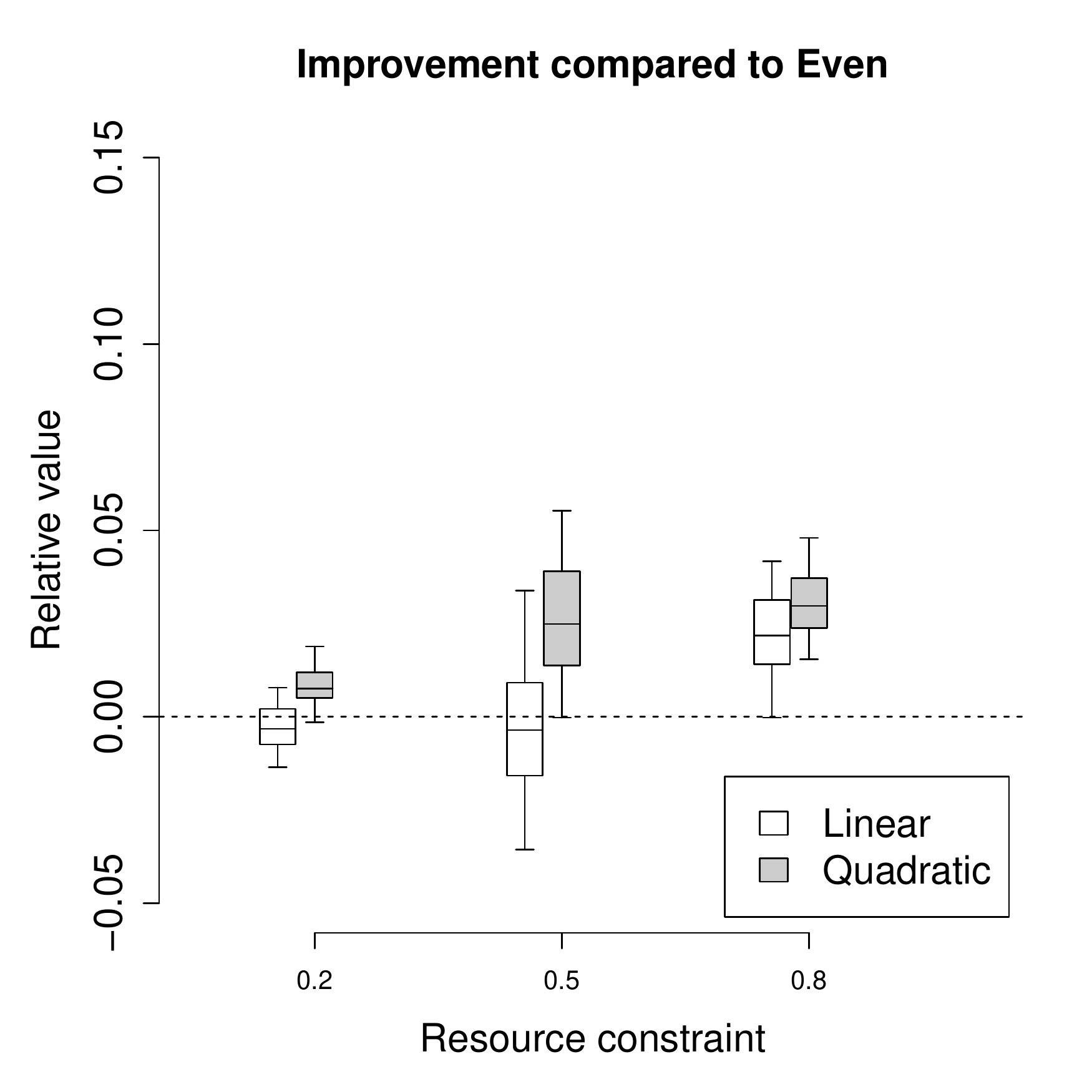}
			\end{picture}
		\end{center}
	\end{subfigure}
	
\end{figure}

We use the two naive policies ``Highest\_rate" and ``Even" as two baseline policies and show the improvement of our proposed policies in terms of loss value compared with the baseline policy. For each of the simulated dataset, we compute the improvement as $(L_{hr}^i-L_{l}^i)/L_{hr}^i$,  $(L_{ev}^i-L_{l}^i)/L_{ev}^i$ and  $(L_{hr}^i-L_{q}^i)/L_{hr}^i$,  $(L_{ev}^i-L_{q}^i)/L_{ev}^i$. Figure \ref{f:imprv}(a) plots the sampling distribution of the improvement of our proposed policies with different utility functions. Under this simulation setting, ``Highest\_rate'' policy is preferred over ``Even'' policy. But we can see our proposed policies have significant improvement compared with either of the naive policies, especially when there are moderate level of total resources ($\mathcal{C}=0.5$). The policy with the linear utility function works slightly better than the policy with quadratic utility function when the total resource level is low ($\mathcal{C}=0.2$). This indicates more extreme resource allocation might improve the overall benefits under this specific simulation setting. 
We conduct another simulation assuming the disease transmission model is misspecified to check the robustness of our method. We assume the true model is more complicated than our proposed model by adding a quadratic effect of the resource allocation on the disease progression, i.e.
$$\boldeta_{t} =\bW_t\boldeta_{t-1}+c_{0}+b_0\bA_t+d_0\bA_t\circ\bA_t+\sum_{k=1}^{p}\beta_{1k}\bX_{k}+\sum_{k=1}^{p}\beta_{2k}(\bX_{k}\circ\bA_{t})+\bepsilon_{t}.$$
We simulate the disease spread for $T=5$ years with the latent process
\begin{multline*}
\eta_{lt}=0.2-0.8A_{lt}+0.2A_{lt}^2+(0.9-0.1A_{lt})\eta_{lt-1}+(0.1-0.1A_{lt})/m_l\sum_{j\in \mathcal{I}_l }\eta_{jt-1}\\+0.12X_{1l}-0.1X_{1l}A_{lt}+\epsilon_{lt},
\end{multline*}
and all other settings are the same as the simulation above.
We can see from Figure \ref{f:imprv}(b) that although our proposed model simplified the true disease spread model, our proposed policies still perform significantly better than naive policies for most cases. Under this simulation setting, the policy with the quadratic utility function works much better than the policy with the linear utility function as this simulation setting prefers more even resource allocation. 

From the simulation results, we can see under different simulation settings and even when our proposed disease progression model is misspecified, our proposed policies are significantly better than naive policies and also consider fairness by allocating the resources more smoothly.

\section{Application to the DRC data} \label{s:real}
We illustrate our method using data in the Democratic Republic of the Congo (DRC) primarily based on Demographic Health Surveys (DHS). DHS are cross-sectional, population-based cluster household surveys. In each survey, clusters are randomly chosen to representative of the national population. Within each cluster, households are randomly selected to participate in the survey. Two DHS program surveys - one in 2007 and another in 2014 - were conducted to study malaria prevalence and treatment allocations in DRC. In the survey, structured questionnaires are administered to selected households to collect malaria-related information, such as their treatment status including bednet use. Also, dried blood spots were collected to test the malaria status. 

\cite{bhatt2015effect} made use of the data from the DHS program surveys and five additional non-DHS program surveys and built a Bayesian hierarchical model to construct a malaria endemicity map across African from 2000 to 2015 in terms of Plasmodium falciparum parasite rate (PfPR). Interventions coverage levels from 2000 to 2015, including insecticide treated nets (ITN), indoor residual spraying (IRS) and Artemisinin based combination therapy (ACT), are also estimated. We download the surface data of PfPR and ITN for DRC from \url{https://map.ox.ac.uk/country-profiles/#!/COD}. In the surface data, PfPR and ITN rate are estimated at a 5km by 5km resolution. Using the smoothed surfaces as data may bias our parameter estimates, but they greatly expand the spatiotemporal coverage of our data which is needed to build a disease progression model. 

As malaria intervention resources are allocated in health zone level in DRC, we map the surface data of PfPR ($Y_{lt}$) and ITN coverage rate ($A_{lt}$) to each of the 515 health zones in DRC by taking the average values of the small cells lying in each health zone as the value of PfPR or ITN coverage rate in the corresponding health zone. We make the assumption that the populations at each 5km by 5km cell within a health zone are the same so that the mean rate of the cells can be used as the rate of the health zone. As there are almost no ITN usage before 2007, we only use the mapped data from 2007 to 2015 to train the Bayesian spatiotemporal model. In the model, we include two environmental covariates ($\bX_1$ and $\bX_2$): annual average temperature and annual average precipitation. As the annual average temperature and annual average precipitation are relatively stable in a certain area, we assume the annual average temperature and annual average precipitation are constant over time in each health zone. We download the monthly worldwide average temperature and precipitation data for 1970-2000 from \url{http://worldclim.org/version2} and map them to health zone level. We use the standardized mean annual average temperature and precipitation in 1970-2000 as the constant environment covariate values used in the model training and prediction.

The collected and processed data are fitted to the model in (\ref{measure}), (\ref{process_uni}) and (\ref{error}). We use 5,000 iterations in Gibbs sampling and discard first burn-in 2,000 samples to obtain 3,000 posterior samples. Table \ref{t:fit} summarizes the posterior mean and 95\% credible interval of the parameters in the model. Temperature, previous disease status and previous neighborhood disease status all have significant effects on the disease spread and the intervention ITN can significantly decrease the disease rate. Environmental covariates and previous disease status also have interaction effects with ITN on the disease progression. 
\begin{table}
	\centering
	\caption{\small{The posterior mean and 95\% credible interval for parameters. The posterior mean with ``$\star$'' represents the corresponding $95\%$ credible intervals that excludes zero.}}
	\label{t:fit}
	\begin{tabular}{lcclcc}
		\hline
		Response& Mean & 95\% CI & & Mean & 95\% CI \\
		\hline
		Intercept ($c_0$) & -0.131$^\star$ & (-0.192, -0.070) &$\sigma_e^2$ & 0.011$^\star$ & (0.010, 0.012) \\
		ITN ($b_0$) & -0.302$^\star$ & (-0.366, -0.234) & $\sigma_2^2$ & 0.142$^\star$ & (0.139, 0.146) \\
		Temperature ($\beta_{11}$) & 0.033$^\star$ & (0.020, 0.045) & $\rho$ & 0.999$^\star$ & (0.998, 0.999)  \\
		Precipitation ($\beta_{12}$) & 0.003 & (-0.012, 0.018) &  & & \\
		Temperature*ITN ($\beta_{21}$) & -0.097$^\star$ & (-0.125, -0.069) &  &  & \\
		Precipitation*ITN($\beta_{22}$) & -0.053$^\star$ & (-0.088, -0.019) &  & & \\
		Previous ($1+c_1$) & 0.917$^\star$ & (0.902, 0.932) & &  & \\
		Previous*ITN ($b_1$) & 0.096$^\star$ & (0.062, 0.129) & &  & \\
		Prev\_neighbor ($c_2$) & 0.040$^\star$ & (0.016, 0.065) & & & \\
		Prev\_neighbor*ITN ($b_2$) & -0.092$^\star$ & (-0.145, -0.039) & &  & \\
		\hline
	\end{tabular}
\end{table}

We consider four risk factors in the priority score in the policy: the standardized annual average temperature ($f_{1lt}=X_{l1}$) and the standardized annual average precipitation ($f_{2lt}=X_{l2}$,), logit of disease rate at previous time point $f_{3lt}=Y_{lt-1}$ and mean logit of disease rate of neighborhood zones at previous time point $f_{4lt}=(1/m_l)\sum_{j\sim l}Y_{jt-1}$. We evaluate the policies with either the linear local utility function or the quadratic local utility function and use the resource constraint level $\mathcal{C}=0.5$. 

\begin{figure}
	\caption{\label{fig:pos}Posterior distribution (5th, 25th, 50th, 75th, 95th percentiles) of the weights $\balpha_{opt}$ for risk factors and the spatial penalty term when using linear utility function (left) or quadratic utility function (right). The risk factors include temperature (standardized), precipitation (standardized), current disease rate (logit), current neighborhood disease rate (logit).}
	\begin{subfigure}{0.49\textwidth}
		\label{fig:pos_linear}
		\caption{Linear utility function}
		\includegraphics[width=1\linewidth]{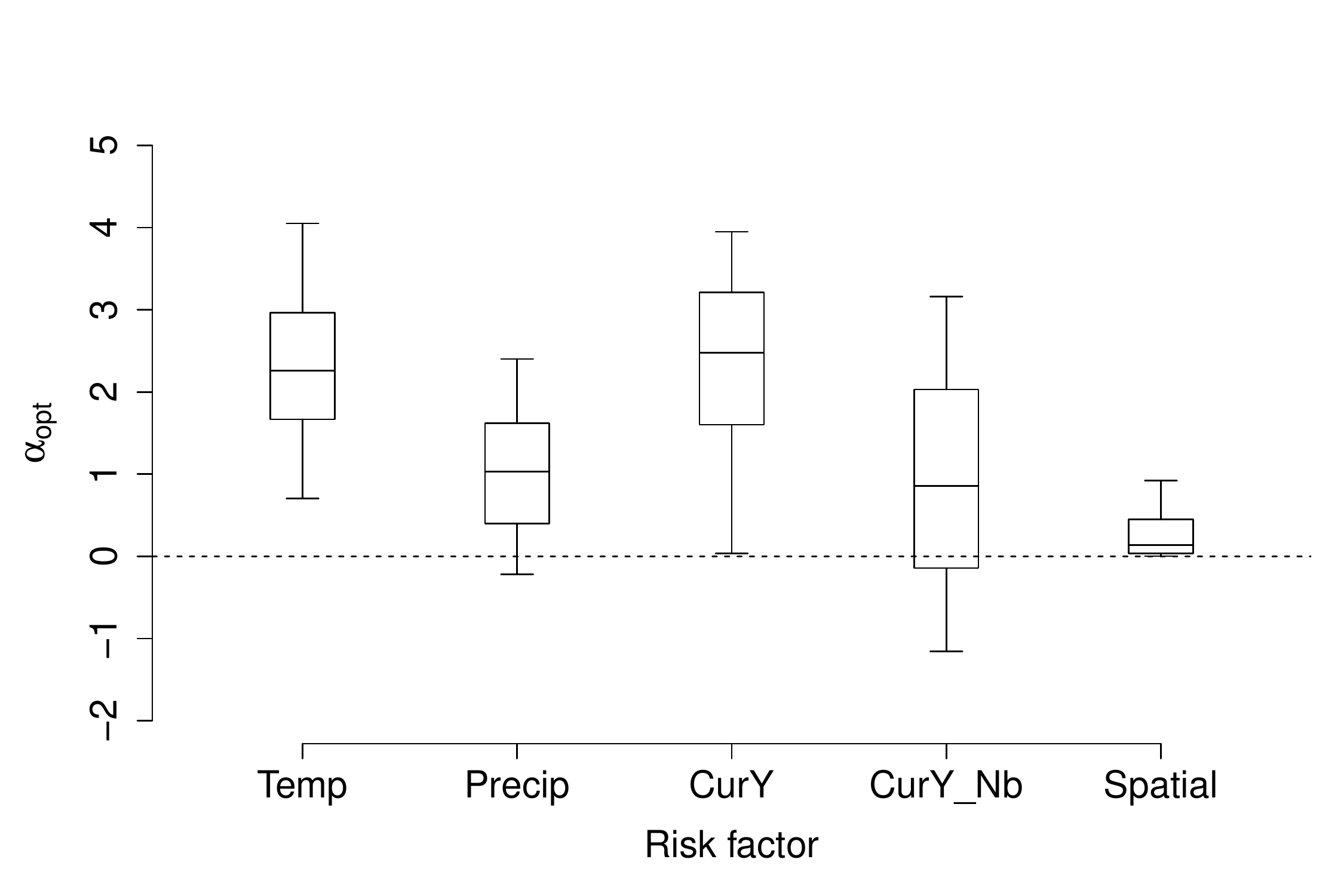}
	\end{subfigure}
	\begin{subfigure}{0.49\textwidth}
		\label{fig:pos_quad}
		\caption{Quadratic utility function}
		\includegraphics[width=1\linewidth]{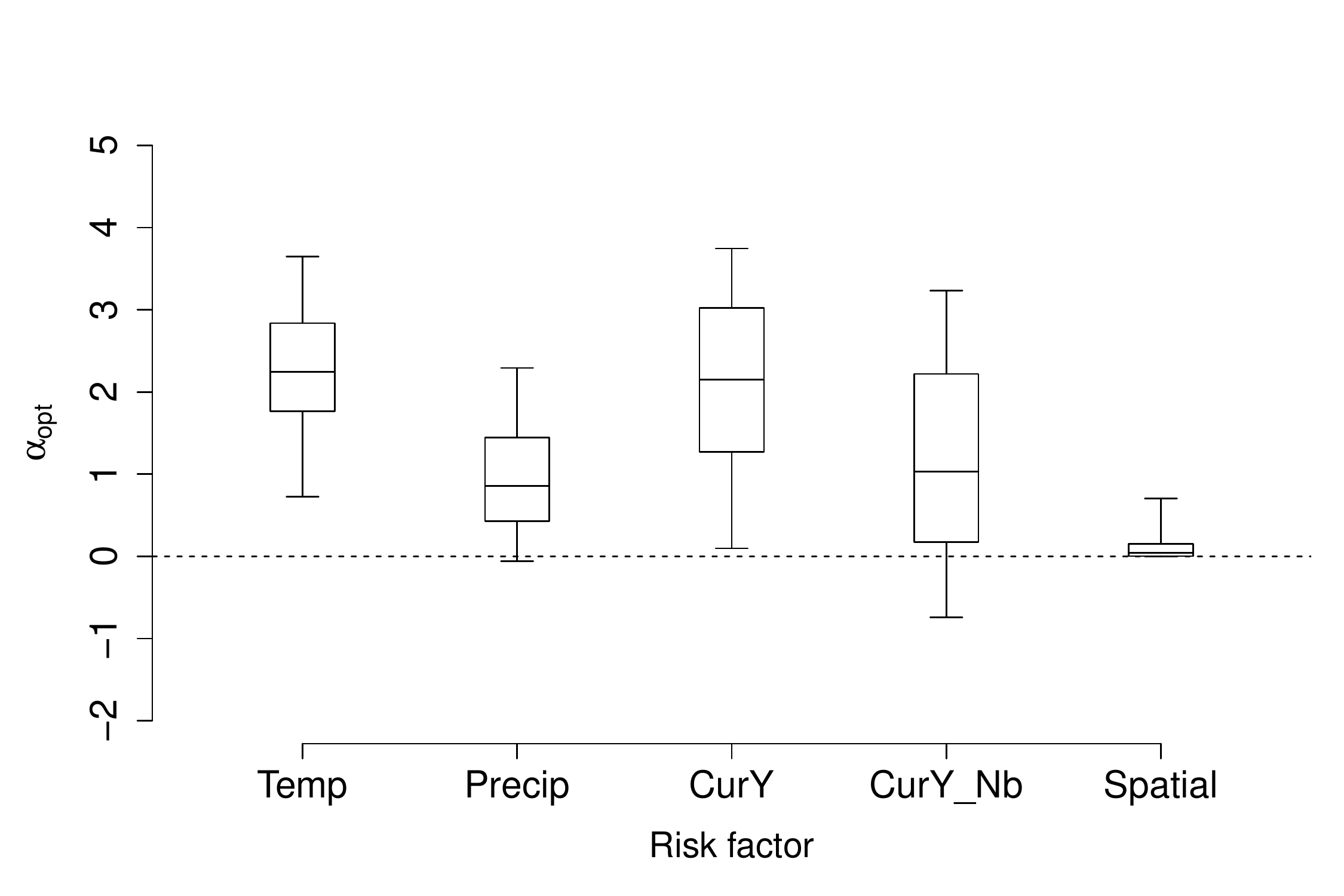}
	\end{subfigure}
	
\end{figure}

We randomly draw 100 posterior samples of the parameters and estimate the optimal policy in terms of $\balpha_{opt}$ corresponding to each posterior draw to get the posterior distribution of $\balpha_{opt}$. Unlike typical MCMC sampling which suffers from autocorrelation, the samples of $\balpha_{opt}$ should be independent draws from the posterior and so fewer samples are needed for these parameters than are needed in typical MCMC sampling. Figure \ref{fig:pos} plots the posterior distribution of $\balpha_{opt}$ when using either the linear utility function or the quadratic utility function. The posterior mean weights for all risk factors are positive, which suggests the priority of being allocated resources for each health zone tends to be positively correlated with temperature, precipitation, current disease status and current neighborhood disease status. For both utility functions, temperature and current disease rate of the health zone seem to be most important factors in determining the risk score and priority. The posterior distribution of weight for the spatial penalty term concentrates more towards to zero when using the quadratic utility function compared with using the linear utility function. This suggests that the spatial penalty term does help to better allocate the resources in terms of smoothness and efficiency when the linear utility function is used, but does little to help when the quadratic utility function is used as the quadratic utility function is able to smooth the resource allocation inherently.

We also estimate one optimal policy averaging over the uncertainty of the parameters. The estimated optimal policy with the linear local utility function is with the priority score $$P_{lt}=1/\{1+\exp[-(2.1X_{l1}+1.3X_{l2}+3.1Y_{lt-1}+0.77\sum_{j\sim l}Y_{jt-1}/m_l)]\}$$ and the weight for the spatial penalty $\alpha_0=0.06$. The estimated optimal policy with the quadratic local utility function is with the priority score $$P_{lt}=1/\{1+\exp[-(3.5X_{l1}+1.1X_{l2}+3.3Y_{lt-1}+0.23\sum_{j\sim l}Y_{jt-1})/m_l]\}$$ and the weight for the spatial penalty $\alpha_0=0.03$.  The estimated optimal weights suggest that health zones with higher temperature, more precipitation, higher disease rate in the zone and neighborhood zones tend to have higher priority to be allocated more resources. The loss value corresponding to the two optimal polices are 0.135 and 0.136 respectively and the loss values corresponding to ``Highest\_rate" policy and ``Even" policy are 0.140 and 0.149, all with standard error $0.0005$. The proposed policies improve the value by about $3\%$ and $9\%$ compared to the two naive policies. This is a substantial improvement when considering the number of disease cases that can be eliminated. 

\begin{figure}
	\caption{\small{The resource allocation next year using the estimated optimal policies with either the linear local utility function (left) or the quadratic local utility function (middle), or using ``Highest\_rate" policy(right).}}\label{f:allocation}
	
	\includegraphics[width=0.32\textwidth]{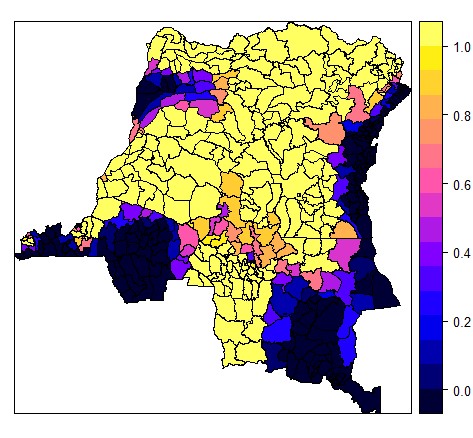}
	\includegraphics[width=0.32\textwidth]{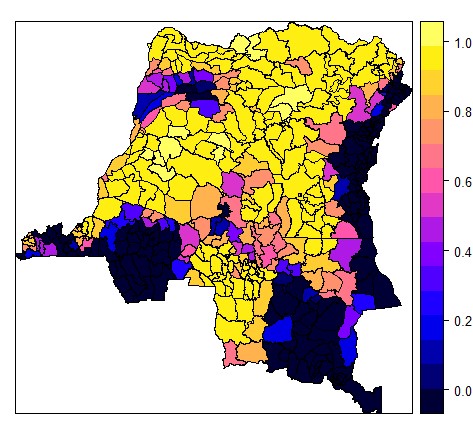}
	\includegraphics[width=0.32\textwidth]{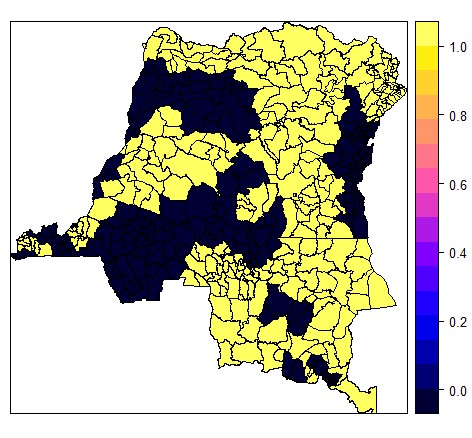}
\end{figure}

Figure \ref{f:allocation} illustrates the resource allocation next year using the two estimated optimal policies or using ``Highest\_rate" rate policy. The estimated optimal policy using the quadratic utility function allocates the resources most smoothly while the ``Highest\_rate" policy only give extreme resource allocation ($0$ or $1$).

\begin{figure}
	\caption{\label{fig:pos_refit}Posterior distribution (5th, 25th, 50th, 75th, 95th percentiles) of the weights $\balpha_{opt}$ for risk factors and the spatial penalty term when using linear utility function and data in year 2007-2012 (left), year 2007-2013 (middle), year 2007-2014 (right) as training data respectively. The risk factors include temperature (standardized), precipitation (standardized), current disease rate (logit), current neighborhood disease rate (logit).}
	\begin{subfigure}{0.33\textwidth}
		\label{fig:pos_linear_07_12}
		\caption{Training data: year 2007-2012}
		\includegraphics[width=1\linewidth]{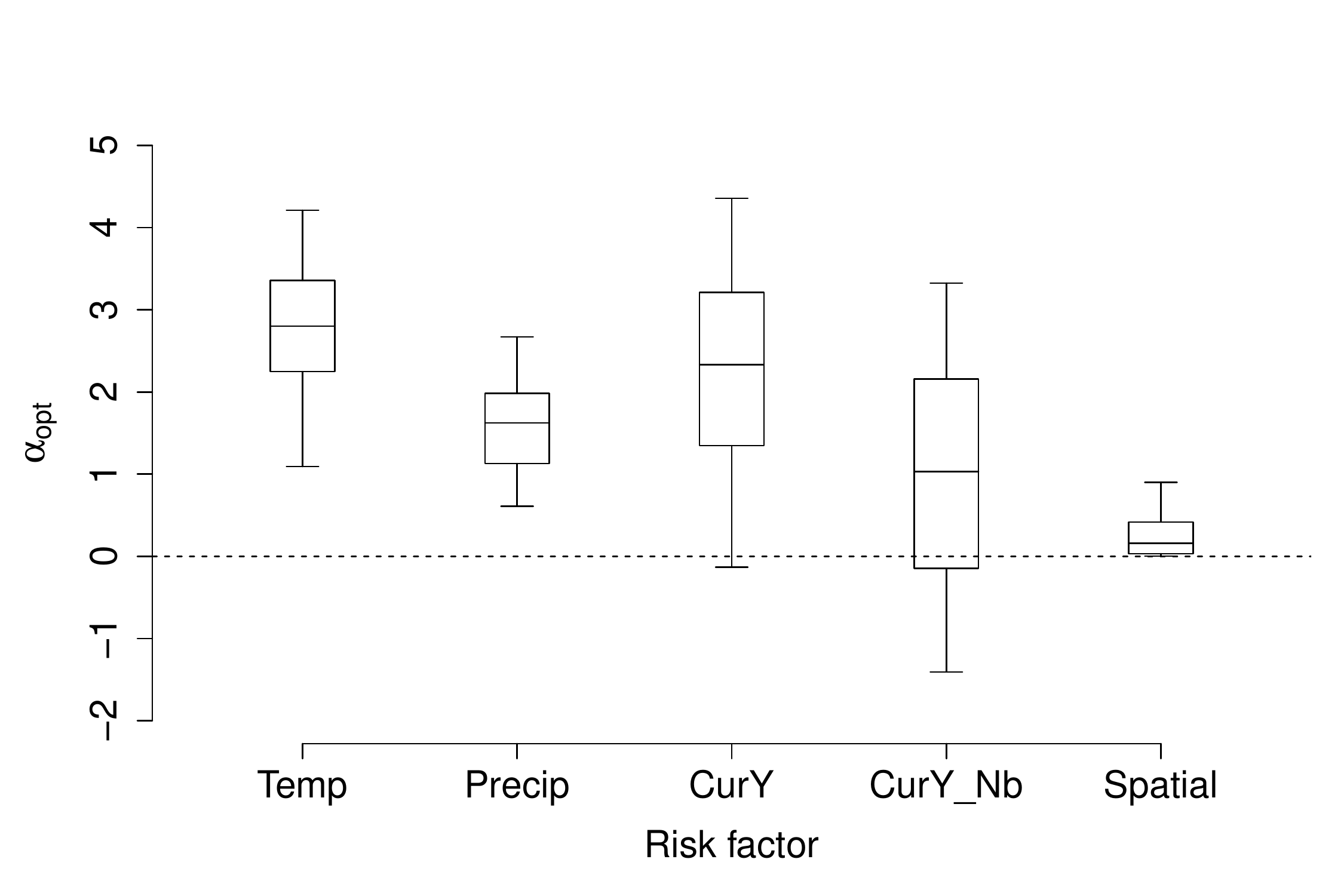}
	\end{subfigure}
	\begin{subfigure}{0.33\textwidth}
		\label{fig:pos_linear_07_13}
		\caption{Training data: year 2007-2013}
		\includegraphics[width=1\linewidth]{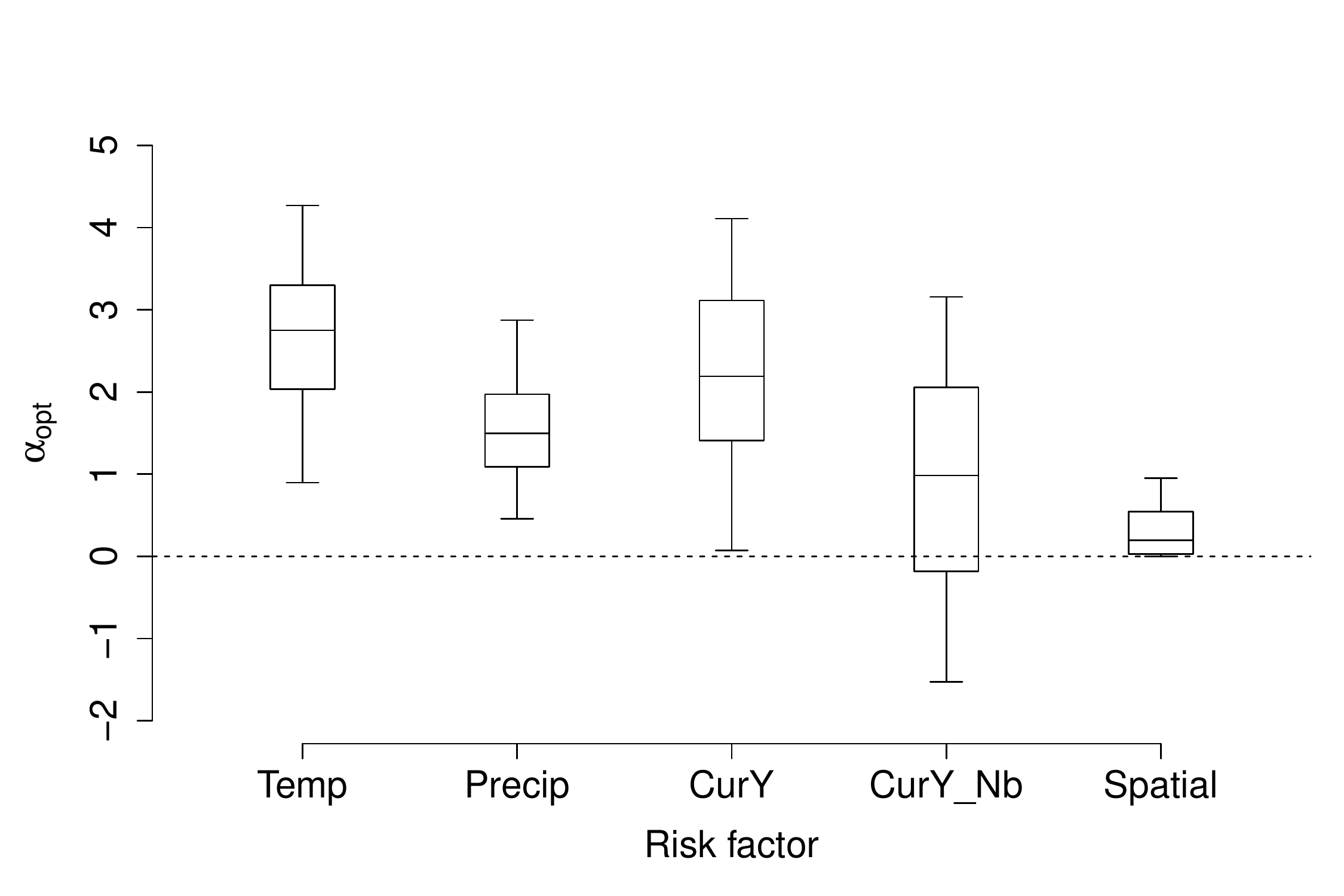}
	\end{subfigure}
	\begin{subfigure}{0.33\textwidth}
		\label{fig:pos_linear_07_13}
		\caption{Training data: year 2007-2014}
		\includegraphics[width=1\linewidth]{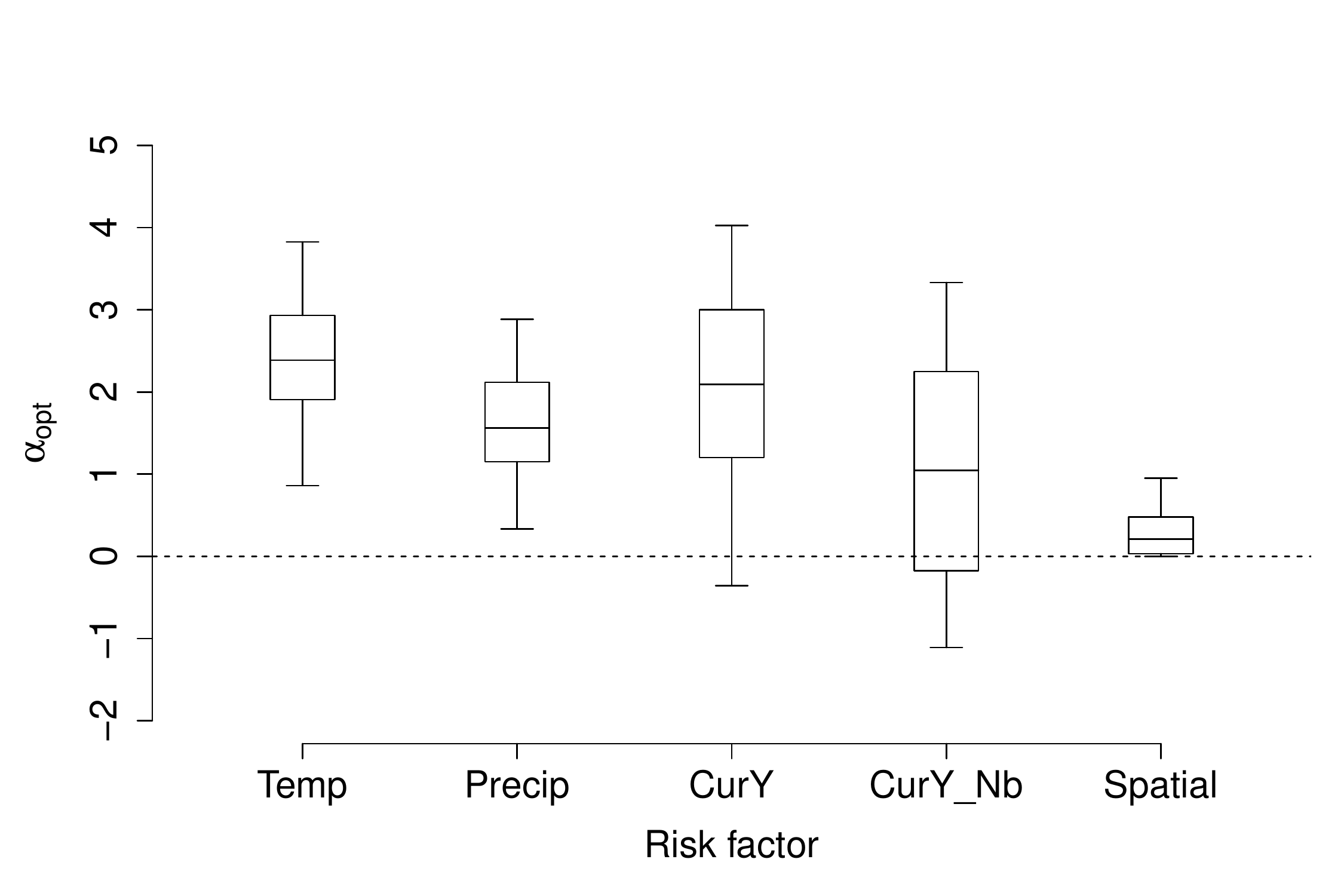}
	\end{subfigure}
	
\end{figure}

We refit the model using data in years 2007-2012, 2007-2013 and 2007-2014 as training data and estimate the risk factor weights $\balpha_{opt}$ with linear utility function for resource allocation recommendation in year 2013, 2014 and 2015 respectively. The posterior distribution of optimal weights are given in Figure \ref{fig:pos_refit}. They show that the updated estimated optimal policy is stable across years suggesting that a dynamic model would not lead to dramatic improvements for this analysis. Of course, for cases where treatment is applied annually there is ample time to update the policy, and so updating the policy using all available data would be advisable.

\section{Discussion} \label{s:discuss}
We develop a recommender system for spatiotemporal resource allocation to maximize the efficacy of malaria control efforts. Our proposed statistical framework deals with the challenges of spatial dependence and continuous action space. We used a hierarchical Bayesian spatiotemporal model to approximate the system dynamics of the disease transmission involving the effect of environmental covariates and allocated resources, and construct a flexible and interpretable class of allocation policies that is also computationally feasible for searching the optimal resource allocation policy with the continuous action space. The simulation experiments suggest the proposed method performs well, and it is shown to be able to improve the resource allocation efficacy compared with naive polices in both simulation studies and the application to DRC data. 

There are some limitations in our studies which provide possible directions for the future work. We make the simplifying assumption that the decision maker will completely comply with the recommended allocation when evaluating the future potential outcomes for each regime. If partial compliance is possible, a compliance model for the distribution of the decision maker's actual action conditioning on the recommended action can be added to the current framework. We assume that a stream of malaria prevalence data for each health zone and each year is available, however, realisticly only partial data may be available due to the difficulty of continuous survellience. To deal with this problem, our proposed Bayesian model could easily inpute the missing data conditional on all the observed data. Our proposed recommendation engine relies on the postulated spatiotemporal model for the malaria transmission. A more flexible semi- or non-parametric model can be constructed to improve the robustness of the method to model misspecification. In the current framework, we only consider one intervention (ITN) at a time when optimizing the resource allocation. Our method can be extended to consider several interventions simultaneously and recommend the optimum allocation policy for all the related resources. As malaria intervention resources are allocated at health zone level in DRC, we build the spatiotemporal model based on the use of data aggregated at the district level and define recommendation engine to decide how to allocate resources to a finite number of regions. If applying treatment to points rather than a finite number of regions is more of interest and data are available at point locations, then a geostatistical model would be preferable to avoid bias in estimating covariate effects and a completely new framework is required to define the point level treatment allocation policy, and it is a topic for future work.

When we apply our method to the DRC data, the data we use to fit the model is the generated PfPR and ITN surface data estimated in \cite{bhatt2015effect} using the Bayesian hierarchical model based on very sparse survey data instead of the real yearly collected health zone level data. Also, we make the assumption that the populations at each 5km by 5km cell within a health zone are the same so that the mean rate of the cells can be used as the rate of the health zone. As a result, we only use the data as the illustration purpose instead of accurately reflecting the situation in DRC.

 \begin{singlespace}
 	\bibliographystyle{biom}
 	\bibliography{malaria}
 \end{singlespace}
 
\end{document}